\documentclass[twocolumn]{article}
\usepackage{amssymb,amsfonts,amsthm,bm,graphicx,amsmath,float,color,longtable}
\restylefloat{figure}
\usepackage{authblk}
\usepackage{pdfpages}

\usepackage{amsmath,amssymb,amsfonts,upref,endnotes}
\usepackage{amsthm}

\usepackage{amsmath,amsthm,amssymb,amsbsy,epsfig,fancyhdr,calc,ifthen,float,psfrag,mathrsfs,graphics,color,fullpage}

\title{The Limiting Speed of the Bacterial Flagellar Motor}

\author[1]{Jasmine A. Nirody}
 \affil[1]{Biophysics Graduate Group, University of California, Berkeley, Berkeley, CA 94720 USA}
\author[2]{Richard M. Berry}%
\affil[2]{Department of Physics, University of Oxford, Oxford OX1 3PU United Kingdom}

\author[3]{George Oster}
\affil[3]{Department of Molecular and Cellular Biology, University of California, Berkeley, Berkeley, CA 94720 USA}

\date{\today}

\begin{document}
 \twocolumn[
  \begin{@twocolumnfalse}
\maketitle
\begin{abstract}
Recent experiments on the bacterial flagellar motor have shown that the structure of this nanomachine, which drives locomotion in a wide range of bacterial species, is more dynamic than previously believed. Specifically, the number of active torque-generating units (stators) was shown to vary across applied loads. This finding invalidates the experimental evidence reporting that limiting (zero-torque) speed is independent of the number of active stators. Here, we propose that, contrary to previous assumptions, the maximum speed of the motor is \textit{not} universal, but rather increases as additional torque-generators are recruited. This result arises from our assumption that stators disengage from the motor for a significant portion of their mechanochemical cycles at low-loads. We show that this assumption is consistent with current experimental evidence and consolidate our predictions with arguments that a processive motor must have a high duty ratio at high loads.
\end{abstract}
  \end{@twocolumnfalse}]

The bacterial flagellar motor (BFM) drives swimming in a wide variety of bacterial species, making it crucial for several fundamental processes including chemotaxis and community formation \cite{Berg2003,korobkova2006hidden,Bai2010,sourjik2012responding}. Accordingly, gaining a mechanistic understanding of this motor's function has been a fundamental challenge in biophysics. The relative ease with which the output of a single motor can be measured in real time, by observing with light microscopy rotation of a large label attached to the motor, has made it one of the best studied of all large biological molecular machines. 
\begin{figure}
\begin{center}
\includegraphics[width=0.4\textwidth]{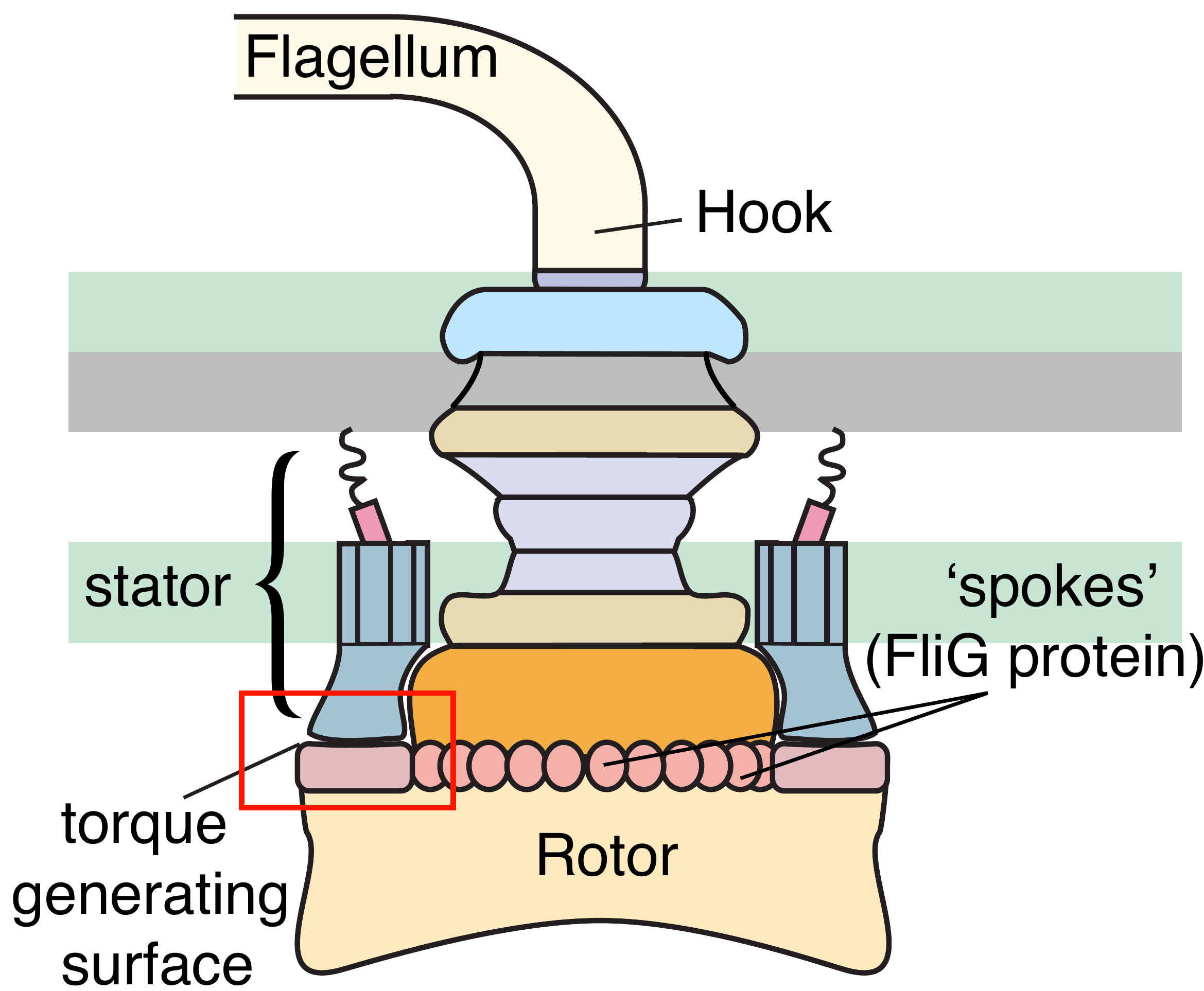}
\caption{The bacterial flagellar motor consists of a series of large concentric rings that attach to a flagellar filament via a flexible hook. An active motor can have between 1 and 11 torque-generating stator units. Stators interact with protein `spokes' (FliG) along the rotor's edge to drive motor rotation.}
\label{schematic}
\end{center}
\end{figure}
However, because of its complexity and localization to the membrane, atomic structures of the entire motor are not yet available. Still, relatively detailed models have been developed using a combination of partial crystal structures \cite{Lloyd1999,Brown2002,Lee2010}, cross-linking and mutagenesis \cite{Zhou1998,braun2004arrangement,lowder2005flig}, and electron microscopic and cryo-electron tomography images \cite{khan1992cytoplasmic,suzuki2004structure} (Fig~\ref{schematic}).

\begin{figure}[t!]
\includegraphics[width=0.49\textwidth]{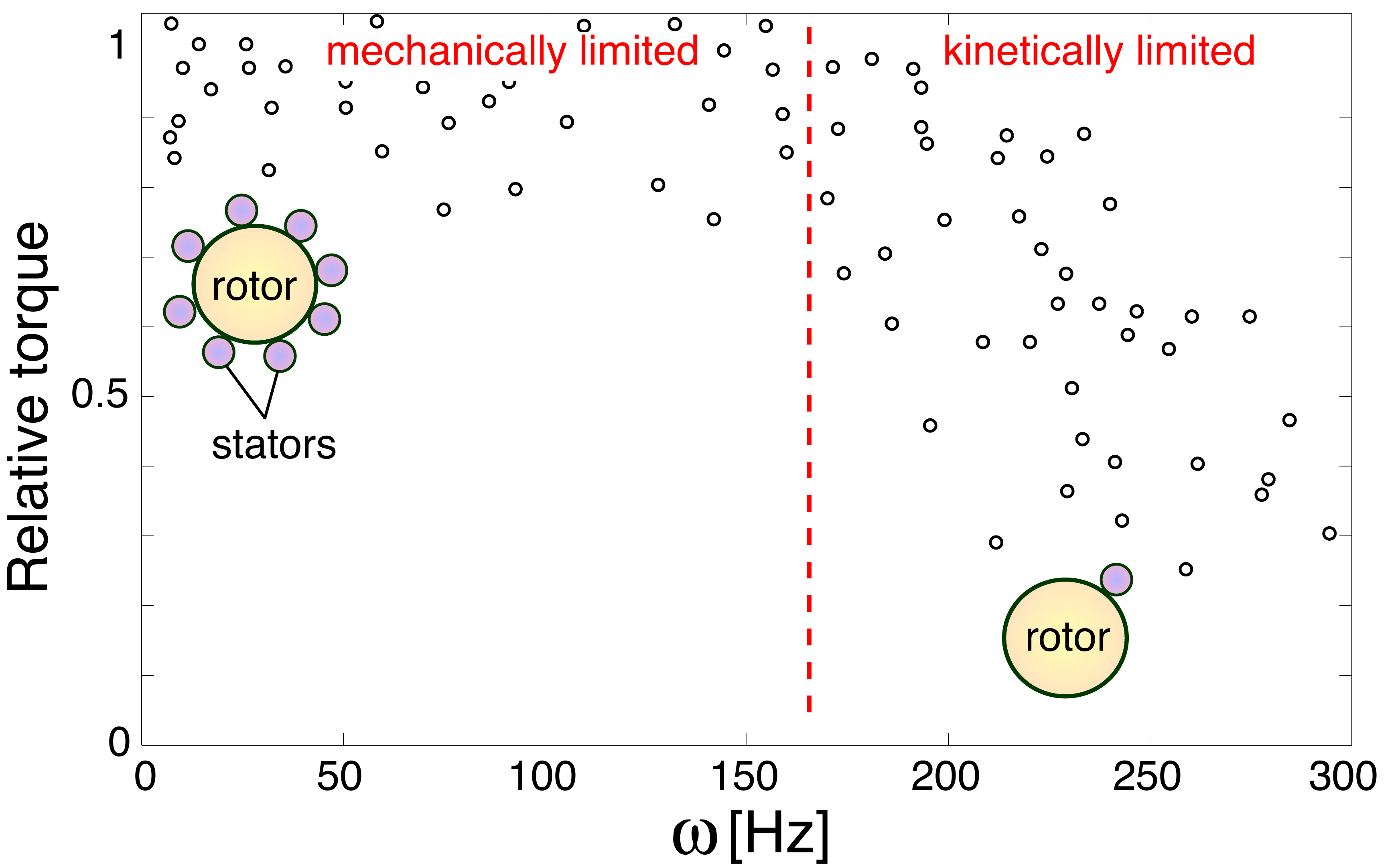}
\caption{Recent experiments have shown that the number of torque-generators (stators) is not constant across applied loads. Therefore, it is likely that previously measured torque-speed curves (here, data from \cite{fung1995powering}) were generated using motors with varying numbers of stators: points in the high-load regime correspond to motors with up to 11 stators and points at low loads to motors with only one. Red dashed line separates the high-load, mechanically-limited and low-load, kinetically-limited regimes of the curve; the latter is the focus of this article.}
\label{intro}
\end{figure}

Arguably the most important physical probe into the \textit{dynamics} of a molecular motor is its torque-speed relationship. For the BFM, this curve was shown to have two distinct regimes (Fig.~\ref{intro}). This characteristic feature of the BFM was long held as the first `checkpoint' for any theoretical model of the motor. However, recent experiments showed that the number of torque-generating units (\textit{stators}) in the motor is load-dependent---that is, published torque-speed curves most likely contain measurements from motors with different numbers of engaged stators \cite{Lele2013,tipping2013load}. Specifically, at high loads (low speeds) a motor can have up to 11 engaged stators, while at low loads (high speeds) motors typically operate with only one stator. 

This finding shed doubt on several fundamental results about the dynamics of the BFM, including, importantly, its behavior at low loads. A seminal set of experiments, termed `resurrection' experiments, studied the dependence of motor speed on the number of stators at various external loads \cite{block1984successive,Reid2006,Yuan2008}.  In these experiments, paralyzed cells were allowed to begin rotating slowly, and discrete increases in speed were interpreted as the addition of torque-generating units. Surprisingly, while up to 11 increases of near-equal size were observed at high loads, only a single such `jump' was observed at low loads. 

These results quickly led to a series of reworked theoretical models, all of which required that the limiting speed of the motor be independent of the number of torque-generators \cite{meacci2009dynamics,bai2009model,meacci2011dynamics}. However, it is likely that low-load measurements were never performed on motors with more than one stator, leaving open the question of how the BFM behaves in the zero-torque (high-speed) limit. 

Here, we predict that the limiting speed of the BFM increases with the number of active stators. This prediction is due to our assumption that the stator is not in contact with the rotor in between steps, or `power strokes' (i.e., the \textit{duty ratio} of the motor is less than 1). We recently presented a model for torque-generation in flagellar motors with a single stator \cite{mandadapu2015mechanics}. Here, we extend this model to motors with multiple stators. Our model is a specific example of such a mechanism; however, most models involving a conformational change in stator structure will share this property. This is because such mechanisms likely require stators to `reset' between steps. 

We argue that these mechanisms have a significant effect on the motor's duty ratio only at low loads. In this way, our model, and others in this category, remain compatible with current evidence that the BFM must have a high duty ratio to be processive at high loads. Experiments testing this hypothesis, if successful, would be the first to explicitly quantify the behavior of a multi-stator motor in the low-load regime. 

\textbf{Overview.} We have implicated a steric interaction between the stator and rotor in torque generation \cite{mandadapu2015mechanics}. Here, we briefly describe our proposed mechanism. Further details, including explicit forms of the Langevin equations used in simulation, can be found in \cite{mandadapu2015mechanics}, as well as in the supplementary material. 

Stators drive the rotation of the motor by stepping along protein `spokes' around the periphery of the \textit{rotor}, which is a large ring that connects to the flagellar filament via a flexible hook. This interaction is analogous to parents pushing on the handles of a merry-go-round on the playground for their children's amusement.

Individual steps are initiated by the arrival of protons at ion-binding sites within the stator complex. During the power stroke, conformational changes in the stator apply a steric force onto the spokes of the rotor wheel, rotating it a discrete step-length $\ell$. Stators apply no productive torque to the rotor between power strokes. Because the BFM lives at low Reynolds number, the rotor also exhibits no productive movement in between steps.

\begin{figure*}
\begin{center}
\includegraphics[width=0.9\textwidth]{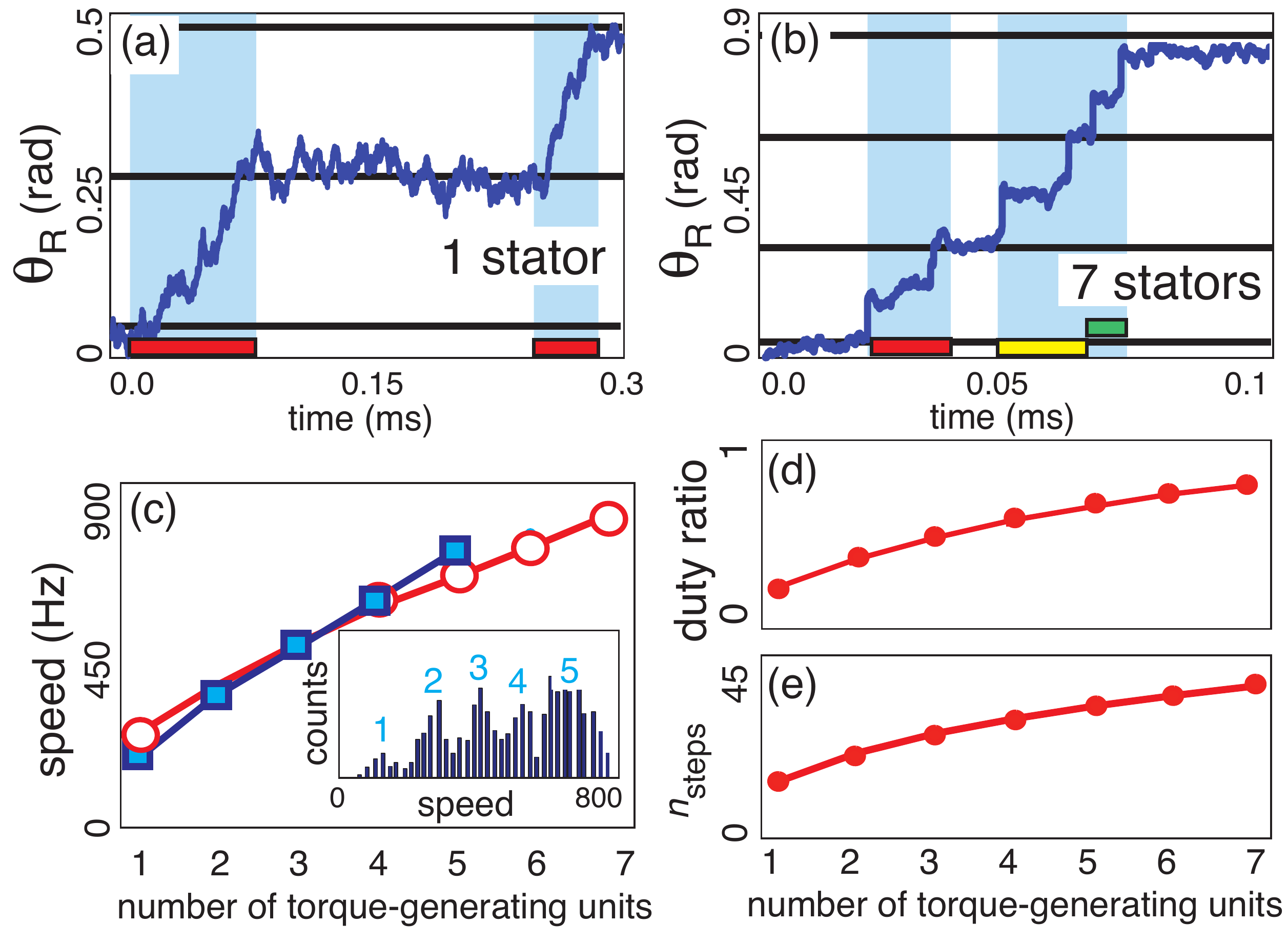}
\caption{Simulated trajectories are shown for motors with \textbf{(a)} one and \textbf{(b)} seven engaged stators, with $\zeta_L = 0.005$ pN-nm-s-rad$^{-1}$. Horizontal black lines denote the distance between ``perfect'' steps ($\ell = \frac{2\pi}{26}$ rad). Colored bars at the bottom of the plots mark the duration of individual stator steps. In the multi-stator motor trajectory \textbf{(b)}, steps for each stator are differently colored. In accordance with published temporal resolutions \cite{nishiyama2002chemomechanical,kolomeisky2007molecular}, we consider individual steps distinguishable if they are separated by 10 $\mu$s. These are shaded in blue; for multi-stator motors, steps may overlap or be too close together to be observed. \textbf{(c)} Motor speed at low loads increases with the number of stators. An experimentally-measured speed distribution at low loads is shown in the inset (data from \cite{lo2013mechanism}). Gaussian fits to the major peaks give mean speeds (blue squares) in good agreement with simulation predictions (open red circles). \textbf{(d)} Decrease in the average time between steps with increasing stator number results in an increase in duty ratio in the low-load regime. \textbf{(e)} As stators are recruited to fast-rotating motors (i.e., at low load), the number of independent stator steps per motor revolution $n_\text{steps}$ increases sublinearly from 26 steps/rev for single-stator motors.}
\label{preds}
\end{center}
\end{figure*}

We assumed that there are 26 spokes along the edge of the rotor (\cite{sowa2005direct}, although see, e.g., \cite{Lee2010,paul2011architecture}). A `perfect' power stroke is defined as a step of length $\ell = \frac{2\pi}{26}$ rad, leaving the stator in contact with the neighboring spoke. These steps are observed through the rotation of a small bead (the `load') attached to a truncated flagellar hook. When the connection between the rotor and the bead is soft, discrete motor steps `blur' into a seemingly continuous trajectory. Experimentally, steps have been directly observed by slowing the motor down to a speed of approximately 10 Hz \cite{sowa2005direct}. Simulation trajectories showing steps for high-speed (near-zero load) motors with one and seven engaged stators are shown in Fig.~\ref{preds}(a)-(b).

\textbf{Motor speed at low loads increases with number of stators.} From simulations, we predict that the maximum speed of the motor is \textit{not} `universal' as currently assumed, but dependent on the number of engaged torque-generators (Fig.~\ref{preds}(c), open red markers). In their recent paper, Lo \emph{et al.} computed torque-speed curves for a chimeric sodium-driven motor \cite{lo2013mechanism}. Low-load measurements on these motors were performed using a 100~nm-diameter gold bead (inset, Fig. \ref{preds}(c)). 

This data was collected from motors with between 1 and 5 active stators, with results from motors with higher stator numbers corresponding to faster peaks (Fig.~\ref{preds}(c), blue markers). The authors chose to focus on the dynamics of single-stator motors, leaving open the implications of this data for how the zero-torque speed depends on stator number. The existence of multiple discrete peaks at low load strongly supports the idea that the maximum speed is dependent on the number of stators, at least in chimeric motors. While experimental results characterizing how the zero-torque speed varies with the number of stators have yet to be published on the wild-type, our predictions should hold for both Na$^+$ and H$^+$ motors.

Previously, Ryu and coauthors reported a set of general conditions that must be met in order for the limiting speed to be independent of the number of engaged stators \cite{ryu2000torque}. First, the rate at which steps are initiated must be independent of the relative position of the rotor and the stator. This position is dependent on both the external load and the actions of any other engaged stators. Therefore, the `decision' of a stator to step should be ignorant of both these factors. Second, stators must engage the rotor for the majority of their cycle (that is, the BFM's duty ratio $DR \approx 1$). Using reasoning based on dynamics at high load, the authors concluded that the duty ratio of the stators was indeed very high. Experiments reporting that the speed at low loads was independent of stator number soon followed \cite{Yuan2008}, which seemed to lend strong support to both of the proposed requirements~\cite{ryu2000torque}.

We assume that the stators are disengaged from the rotor for a large part of their cycle at low loads, resulting in a violation of the second condition. Unlike most recently proposed mechanisms (but see \cite{boschert2015loose}), we assume motor rotation and ion flow can be \textit{loosely coupled}: an ion passage may not always result in appreciable rotation of the rotor. The \textit{stator's} motion, however, is tightly coupled to ion flow---that is, an ion passage is both necessary and sufficient for the initiation of a stator's power stroke. Therefore, loose coupling in our model does not arise from some form of ion leakage, as may be expected~\cite{boschert2015loose,oosawa1982mechanism,oosawa1983coupling,oosawa1986loose,Berry1993}. Instead, it is due to the fact that stator steps are rarely `perfect' in multiple-stator motors: if stator steps overlap, a portion of the second stroke is `wasted' because the rotor is pushed out of the later-firing stator's reach. 

These properties seem contrary to present assumptions that stators in the BFM must have a high duty ratio. However, arguments in support of $DR \approx 1$ are largely based on motor dynamics at high load. We show that our prediction that $DR < 1$ at low loads arises from fundamental differences in motor dynamics between the two regimes. In this way, we argue that our proposed mechanism is compatible with experimental evidence for a high duty ratio at high loads. 

\textbf{Kinetically-limited stators have low duty ratios.} A stator initiates a step when protons arrive at a specified binding site within the complex. The mechanochemical cycle of the stator then has two phases: moving and waiting, characterized by timescales $T_m$ and $T_w$, respectively \cite{meacci2009dynamics}. If $T_S$ is the time that a stator engages the rotor during a complete cycle ($T_m + T_w$), a single-stator motor has duty ratio $DR = T_S/(T_m + T_w)$.

The waiting time between power strokes $T_w$ depends on the rate of proton arrivals at the binding site on a stator unit. These arrivals are Poissonian with rate $k_{\text{on}} = k_0\exp \left[\lambda\Delta G_{\text{ij}}/k_BT\right]$. Here, $\Delta G_{\text{ij}}$ is the thermodynamic contribution of the ion motive force and $k_BT$ is Boltzmann's constant multiplied by temperature \cite{xing2006torque,bai2009model}. For simplicity, we choose $\lambda=0.5$ as done in previous studies \cite{xing2006torque}. The parameter $k_0$ is a function of the pH of the external periplasm; lower pH corresponds to higher proton concentration and thus a speedier arrival at the site. At room temperature and pH 7.0, $\langle T_w \rangle = 1/k_\text{on} =  0.2$ ms for single-stator motors.


The average moving time is estimated through the relation $\omega \approx \ell/\left(\langle T_m \rangle + \langle T_w \rangle\right)$ \cite{meacci2009dynamics}. The average motor speed $\omega$ is also related to the load drag coefficient $\zeta_L$ by $\zeta_L\omega \approx \tau$, where $\tau$ is the motor torque \cite{Berg2003,inoue2008torque}. In our simulations, the motor is limited by proton arrivals at very low loads ($\langle T_m \rangle \approx 0.01$ ms), while at high loads, $\langle T_m \rangle \approx 10$ ms surpasses $\langle T_w \rangle$. These values are consistent with previous studies \cite{meacci2009dynamics,meacci2011dynamics}.  

Because we predict that motor rotation is driven by steric forces, a stator must be in contact with the rotor for a large part of a productive power stroke ($T_S/T_m \approx 1$). Previous models of torque-generation have similarly considered the mechanochemical cycle of the BFM to consist of moving and waiting phases \cite{meacci2009dynamics,meacci2011dynamics}. However, our model is unique in assuming that stators disengage from the rotor between subsequent power strokes. This results in $DR < 1$ for single-stator motors at low loads, as the waiting time is no longer negligible compared to the moving time in this regime (Fig. \ref{preds}(d)). The waiting time may even surpass $\langle T_m \rangle$, as shown in Fig. \ref{preds}(a)-(b). 

The waiting time until a proton binds to any one of $N$ independently-stepping stators is exponentially distributed with rate $N\times k_{\text{on}}$. Therefore, $\langle T_w \rangle$ is shortened as additional stators are recruited. The subsequent increase in duty ratio (Fig. \ref{preds}(d)) results in an increase in limiting speed with the number of stators. 

\textbf{High duty ratios at high loads.} Here, we address two arguments which have been used to assert that the duty ratio of the BFM must be very high: (i) the observation that the number of steps per revolution $n_{\text{steps}}$ increases as additional torque-generating units were recruited \cite{samuel1995fluctuation,samuel1996torque}, and (ii) a calculation determining that a motor with a low duty ratio cannot be processive due to `unwinding' of the tether connection between the rotor and load \cite{Berg2003}. Though these arguments are based on observations at high load, they were taken as support for a zero-torque speed independent of stator number. This extrapolation was possible largely due to the absence of a proposed physical mechanism for rotation of the BFM. Such a mechanism is now provided in our model \cite{mandadapu2015mechanics}. To this end, we show that these arguments can be consolidated with our proposed mechanism, as well as with the corresponding prediction that $DR < 1$ at low loads.

Samuel and Berg used fluctuation analysis to determine that the number of steps per revolution was proportional to stator number \cite{samuel1995fluctuation,samuel1996torque}. In the absence of a specific physical mechanism, this result was interpreted to mean that a motor decreases its elementary step size as it recruits torque generators. This in turn implied a motor with a high duty ratio, in which each unit acts with the $N-1$ others to rotate a fixed distance $d$ \cite{ryu2000torque}.

This observation holds in the high-load (low-speed) regime, which is where these measurements were made. Even though stators disengage between subsequent strokes, the duty ratio of the motor remains very high because the time spent within a power stroke is far greater than the pauses between subsequent strokes ($DR = T_s/(T_m+T_w) \approx T_s/T_m \approx 1$). Furthermore, the rotor is likely always in contact with at least one stator as the steps of individual stators almost certainly overlap. This accounts for the observed proportional increase in $n_{\text{steps}}$ with the number of active stators.

Stator steps still may overlap at low loads (high speeds), though they are less likely to do so because $T_m$ is shorter than at high loads. Our simulations predict that similar analyses in this regime will detect a sublinear increase in $n_\text{steps}$ with stator number (Fig. \ref{preds}(e)).

A second argument for a high duty ratio in the BFM was posed by Howard Berg, who determined that if the BFM did not have a duty ratio of close to unity, it could not be processive \cite{Berg2003}. The reasoning behind this is as follows. Consider an experiment where a cell is tethered to a surface by the hook of its flagella and is spun about by the rotation of the motor at its base. The cell body is large in comparison to the flagellar motor, and accordingly the viscous drag on it is much larger than that on the BFM's rotor. Therefore, if there are no stators to prevent it, the wound tether between the motor and the cell will unwind exponentially: $\theta = \theta_0\exp(-\alpha t)$, where $\theta_0$ is the initial twist and $\alpha$ is the torsional spring constant divided by the rotational drag coefficient of the rotor. A simple calculation showed that unless a motor had a duty ratio of very close to unity, this tether would unwind too quickly for the stator units to keep up.

We note that concrete evidence is still lacking that slowly-rotating tethered motors do not `lose' steps to the tether connection unwinding. Support for tightly-coupled mechanisms came from reports that the number of ions per revolution was directly proportional to motor speed \cite{Meister1987}. However, it was later shown that a loosely-coupled mechanism also produced a linear relationship with the same slope, but non-zero intercept \cite{Berry1993}. Regardless, our model construction and parameter choice is such that the unwinding of the tether does not overwhelm the stator in our simulations (see supplementary information)~\cite{mandadapu2015mechanics}. A final resolution awaits experiments measuring how the ion flux at stall (zero speed) differs between single- and multi-stator motors. 

In contrast to the high-load regime, the relative drags of the bead and the rotor are comparable at low loads. As we approach the zero-torque limit, the rotor drag may surpass that of the load \cite{meacci2009dynamics,meacci2011dynamics}. For example, we estimated the drag coefficient for the low-load measurement in \cite{lo2013mechanism} to be $\zeta_L \approx 0.005 $ pN-nm-s-rad$^{-1}$, which is lower than $\zeta_R \approx 0.02$ pN-nm-s-rad$^{-1}$ \cite{Berg2003}. In this case, the bead will move forward as the tether connection unwinds.

More generally, the characteristic timescale of the load's motion is given by its frictional drag coefficient divided by the spring constant: $t_L = \zeta_L/\kappa$. A single-stator motor should have a  power stroke of comparable length. Note that this is not necessary for a multi-stator motor: steps from different stators may overlap, extending the period during which at least one unit is engaged. 

To illustrate, we consider the second-smallest bead used by Lo \emph{et al} \cite{lo2013mechanism}. Estimating $\zeta_L = 0.04 $ pN-nm-s-rad$^{-1}$ and choosing a conservative spring constant $\kappa = 150$ pN-nm-rad$^{-1}$ (at the lower edge of the measured range~\cite{block1989compliance}), the characteristic timescale of the load is $t_L = \zeta_L/\kappa \approx 0.27$~ms. A single-stator motor with this load rotated at $\approx$ 110 Hz \cite{lo2013mechanism}. Recall that motor speed $\omega \approx d/(\langle T_m \rangle + \langle T_w \rangle)$, where the step size $\ell = \frac{1}{26}$ rev and $ \langle T_w \rangle \approx 0.2$ ms. Then $\langle T_m\rangle~\approx \left(\frac{1}{26}\right)/110 - 2$e-4 $\approx 0.15$~ms, which is enough time for the load to (at least partially) `catch up' to the rotor. 
 
\textbf{Conclusions.} The dynamics of the BFM across applied loads have been of great interest since a two-regime torque-speed curve was proposed several decades ago. Recent experiments reporting that the number of stators in a motor varies across loads have opened some interesting questions, and reopened several more. 

For instance, the zero-torque speed has been assumed to be independent of the number of engaged stators based on the results of early `resurrection' experiments \cite{block1984successive,Reid2006,Yuan2008}. Theoretical models after these results were reported have all been constructed to reproduce this behavior at low loads. However, recent experiments strongly suggest that these experiments were never performed on motors with more than a single stator \cite{Lele2013}.


In opposition to current assumptions, our simulations predict that the limiting (zero-torque) speed of the BFM increases with stator number. This relationship arises from our assumption that stators detach from the motor when they pause between steps. This assumption is common to most models in which a conformational change in the stator drives motor rotation. This results in a low duty ratio for motors at low load, where the waiting time between steps is at least on the order of the time spent in a power stroke. Because the power stroke duration is much longer at high loads, the duty ratio in this regime is not affected by this unbound state. In this way, our mechanism is consistent with evidence that processive motors at high load must have a high duty ratio.

Recently, Lo \textit{et al.} presented evidence of increasing zero-torque-speed with stator number in chimeric, sodium-driven motors \cite{lo2013mechanism}. However, this result was not fully explored as the authors focused on understanding single-stator motor dynamics. Further experiments, especially on wild-type motors, would directly test the hypothesis presented here, and be the first to explicitly characterize the low-load behavior of the flagellar motor.

\bibliographystyle{unsrt}
\bibliography{ref.bib}



\section*{Supplementary material}

# Supplementary Material for

# The Limiting Speed of the Bacterial Flagellar Motor

Jasmine A. Nirody, Richard M. Berry, George Oster[†]
[†]Corresponding author. Email: goster@berkeley.edu.

# Contents

# 1 Model for torque generation in the BFM

In this section, we provide Langevin equations describing the dynamics of the stator, the rotor, and the load. A more detailed description of the model can be found in [2]. We also detail how the single-stator model can be extended to deal with motors with multiple stators.

## 1.1 Single-stator equations

This model was originally presented and described in detail for single-stator motors in our previous work [2]. We review some important details in this section. The dynamics of the stator, rotor, and load are described by the following Langevin equations:

$$\textbf{Stator}: \quad \zeta_S \frac{d\phi_S}{dt} = \underbrace{F_p \ell_p}_{\substack{\text{Torque from} \\ \text{Proline hinge}}} - \underbrace{\tau_\text{reaction}}_{\substack{\text{Reaction} \\ \text{from rotor}}} + \underbrace{\sqrt{2k_B T \zeta_S} f_n(t)}_{\text{Thermal fluctuations}} \quad (1)$$

$$\textbf{Rotor}: \quad \zeta_R \frac{d\theta_R}{dt} = \underbrace{\tau_\text{contact}}_{\substack{\text{Torque} \\ \text{from stator}}} - \underbrace{\kappa(\theta_R - \theta_L)}_{\substack{\text{Spring connection} \\ \text{to load}}} + \underbrace{\sqrt{2k_B T \zeta_R} f_n(t)}_{\text{Thermal fluctuations}} \quad (2)$$

$$\textbf{Load}: \quad \zeta_L \frac{d\theta_L}{dt} = \underbrace{\kappa(\theta_R - \theta_L)}_{\substack{\text{Spring connection} \\ \text{to rotor}}} + \underbrace{\sqrt{2k_B T \zeta_L} f_n(t)}_{\text{Thermal fluctuations}}. \quad (3)$$

Here $\zeta_S$, $\zeta_R$, and $\zeta_L$ are the effective drag coefficients of the stator, rotor, and load. The last term in each equation is the stochastic Brownian force, where $f_n(t)$ is uncorrelated white noise.

In Equation (1), the internal torque driving the stator due to the rearrangement of hydrogen bonds caused by a proton binding event is denoted by $F_p \ell_p$. Because the motion of the two halves of the power stroke are mechanically equivalent, we collapse the dynamics of the two loops into a single equation. The contact torque applied to the rotor (in Equation (2)), and consequent reaction torque applied to the stator (in Equation (1)), are given by $\tau_\text{contact}$ and $\tau_\text{reaction}$ respectively.

The rotor and load are connected by a linear spring with constant $\kappa$; the elastic coupling terms in the equations for the rotor and the load thus appear with opposite signs (in Equations (2) and (3), respectively). The elastic constant in the experiments can vary depending on the length of the hook when attaching the bead. In some cases, the hook is cut very short or is stiffened by an antibody linker, which would correspond to a large spring coefficient [6].

The transition rates between the two potentials are given by the rates of protons 'hopping on' and 'hopping off' of the stator binding sites. To satisfy detailed balance when the ion-motive force (IMF) is non-zero, the kinetic coefficients for the reaction are chosen so that

$$\frac{k_\text{on}}{k_\text{off}} = 10^{(\text{pK}_a^p - \text{pH}_\text{periplasm})} \exp\left(\frac{\Delta G_{ij}}{k_B T}\right), \quad (4)$$

where $\Delta G_{ij}$ is the thermodynamic contribution of the IMF (see Figure S1) and $k_B T$ is Boltzmann's constant multiplying temperature. Since $G_1$ and $G_2$ are simply horizontal reflections of one another, $\Delta G_{12} = \Delta G_{21}$. For convenience, we choose the following with $\lambda = 0.5$:

$$k_\text{on} = 10^{-\text{pH}_\text{periplasm}} \exp\left(\lambda\left(\frac{\Delta G_{ij}}{k_B T}\right)\right), \quad (5)$$

$$k_\text{off} = 10^{-\text{pK}_a^p} \exp\left(-(1-\lambda)\left(\frac{\Delta G_{ij}}{k_B T}\right)\right). \quad (6)$$

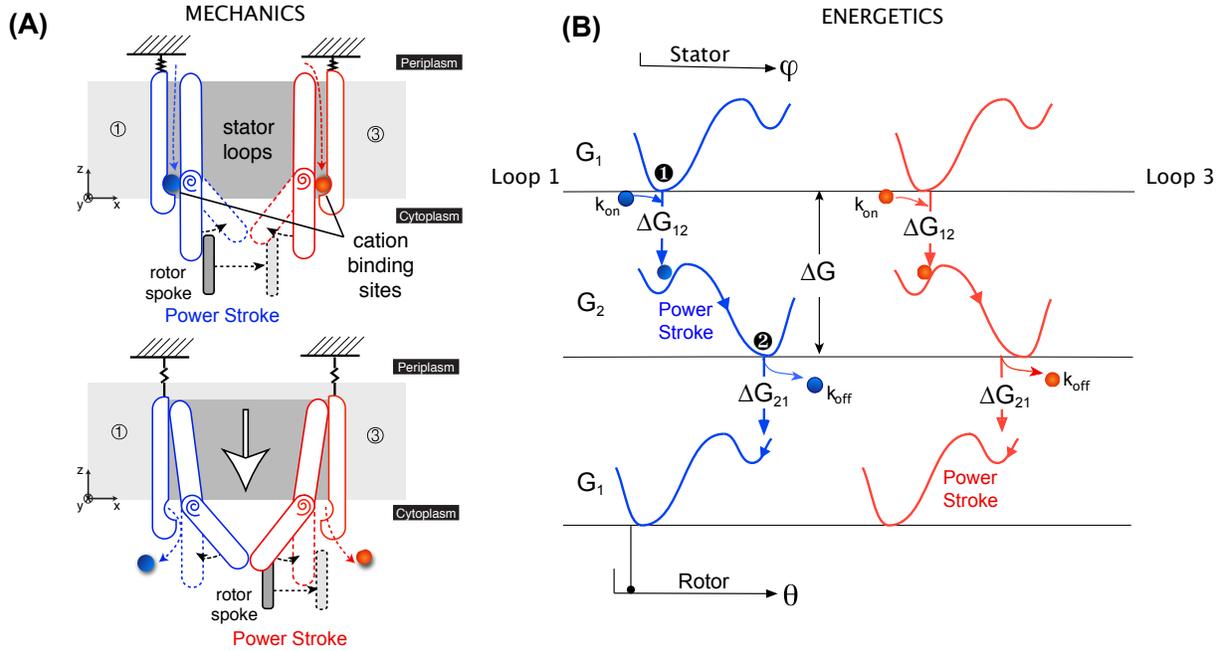

Figure S1: Dynamics of the rotor-stator interaction. **(a)** MECHANICS OF THE POWER STROKE. *Top panel:* Cation binding induces a strain in the stator, which causes the loops to bend. This results in the first half of the power stroke (here, by Loop 1), and sets up the second loop (here, Loop 3) to perform its half of the power stroke. Subsequently, the cations are released into the cytoplasm. This occurs because our proposed motion also has a vertical component—the loops lower themselves out of the membrane. This release then reverses the strain and causes the loops to restraighten. This results in the second half of the power stroke. We note that this image depicts a two-dimensional projection of a three-dimensional motion: the motion of the stators is not constrained to the plane of the page. **(b)** ENERGETICS OF THE POWER STROKE. Because the two loops move in-phase with each other in our model, their energetic pictures are identical. We describe the free energy landscapes using double-well Landau potentials ($G_1$ for the first half of the power stroke, and $G_2$ for the second half). These landscapes are shown in blue for Loop 1 and red for Loop 3 with respect to the angles of the stator $\phi$ and rotor $\theta$. The initial entrance of the proton into the ion channel ($k_{\text{on}}$) places the system within $k_B T$ of the energy barrier. Thermal motions then result in the first half of the power stroke. The exit of the protons into the cytoplasm ($k_{\text{off}}$) results in the 'reset', and the second half of the power stroke. Figure modified from [2].

Table S1: List of parameters used with units, values, and reference.

| Parameter | Definition | Units | Values | Ref |
|---|---|---|---|---|
| $\ell_P$ | Length of the proline hinge arm | nm | 7 | [4] |
| $\zeta_S$ | Drag coefficient of the stator | pN-nm-s-rad$^{-1}$ | 0.002 | fit |
| $\zeta_R$ | Drag coefficient of the rotor | pN-nm-s-rad$^{-1}$ | 0.02 | [3] |
| $\zeta_L$ | Drag coefficient of the load | pN-nm-s-rad$^{-1}$ | 0.005–10 | [5] |
| $\kappa$ | Hook spring constant | pN-nm-rad$^{-1}$ | 150 | [6] |
| $N$ | Number of stators | - | 1–11 | [5] |
| $\phi_S$ | Angular position of the stator | rad | - | - |
| $\theta_R$ | Angular position of the rotor | rad | - | - |
| $\theta_L$ | Angular position of the load | rad | - | - |

## 1.2 Extension to multiple-stator motors

In a motor with multiple stators, the mechanics of each stator are as described above. In this section, we describe how we can extend the above model to deal with multiple independently-stepping stators. The mechanics of each unit follows the equations presented for a single stator. In particular, each stator is independently 'activated' at rates given by Equations (5) and (6). Because cation arrivals are Poisson processes (i.e., waiting times between arrivals are distributed exponentially) [7, 8], the 'next arrival' in a motor with $N$ stators occurs at a rate $N \times k_{\text{on}}$, where $k_{\text{on}}$ is the rate of arrival for a single stator.

An important consideration in simulations with multiple, independently-stepping stators is the following. While each stator pushes on its own 'spoke' on the rotor's edge, these spokes are rigidly connected. This means that if one stator begins its power stroke shortly after another stator has done so, it likely will apply no torque to the rotor for some portion of its cycle. This is because the positions of the rotor spokes are dependent on each other, and the power stroke of the first stator will have pushed the second stator's spoke slightly out of reach (at least for the initial part of its cycle).

## 2 Numerical implementation

In this section, we discuss briefly some technical details regarding the implementation of the above systems of equations. Simulations of Langevin dynamics were written in Python 2.7. Example low-load simulation output can be seen for motors with one and seven stators in Figure S2. Single-stator trajectories at varying loads are shown in Figure S3.

Discrete transitions are modeled using Gillespie's method, as follows. For motors with $N$ stators, $N$ 'first arrival times' are initially chosen from an exponential distribution at $t = 0$. Each subsequent waiting time is drawn from an exponential distribution when the stator loop reached a small range around the potential minima. For example, the time required to 'hop off' is chosen when the angle of the stator loop is within a small range $(20 - \epsilon°, 20 + \epsilon°)$ for some prescribed $\epsilon$. Likewise, the time for the next cation arrival is chosen when the angle retracts to within $\epsilon$ of $0°$. This is done to imitate the alternating access of the cation-binding site to the periplasm and cytoplasm.

The rate for protons hopping off into the cytoplasm ($k_{\text{off}}$) are chosen as 1000 times the value for proton arrivals $k_{\text{on}}$ [9]. This is in line with the fact that half-steps have yet to be directly observed experimentally.

Continuous-time portions of each cycle (corresponding to the mechanical movements) for the stator, rotor, and load are simulated using a forward finite difference scheme with a time step of $10^{-8}$ s. Checks are put in place to assure that the stator position does not surpass the position of the rotor due to the time step being too large.

Rotor spokes (FliG proteins) are rigidly connected to each other. Therefore, if a stator $s_1$ initiates its power stroke at time $t_1$ and a second stator $s_2$ initiates its power stroke at time $t_2 > t_1$, then the FliG

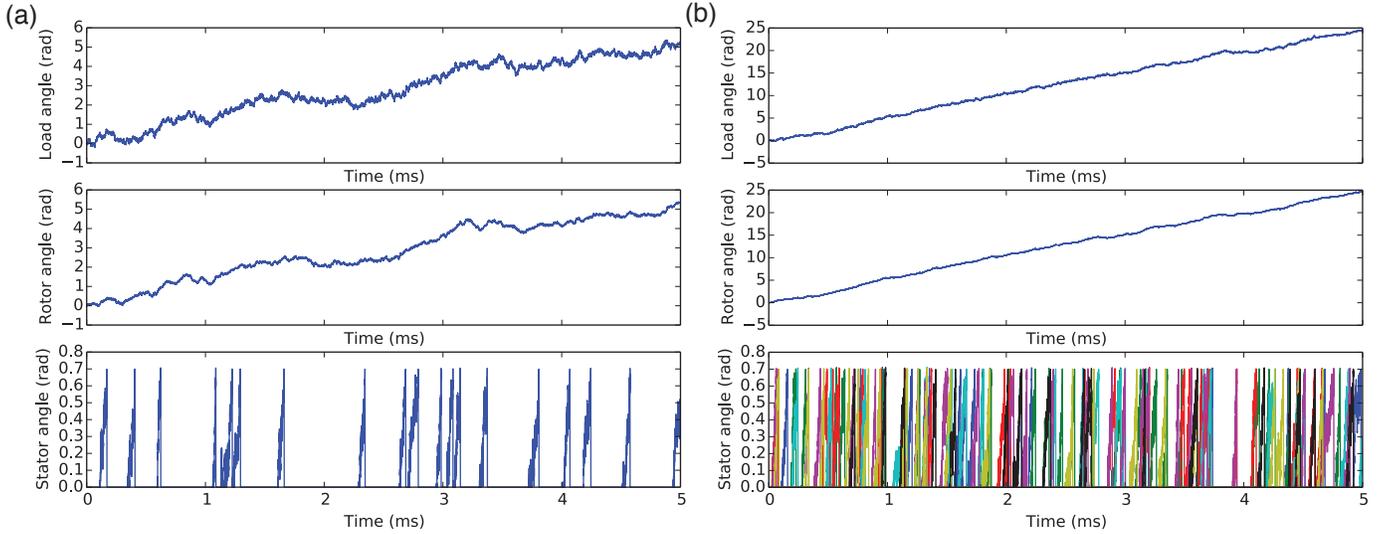

Figure S2: Simulation output (5 ms) at $\zeta_L = 0.005$ pN-nm-s-rad$^{-1}$ for motors with **(a)** one and **(b)** seven stators. Top and middle panels show load and rotor trajectories, respectively. Bottom panels show stator stepping events. In **(b)**, events for each stator are colored uniquely.

adjacent to $s_2$ will have moved the same distance as the FliG that $s_1$ pushed in the interval $[t_1, t_2)$. This means that the portion of the power stroke up until $s_2$ can 'catch up' to the FliG in front of it will be 'wasted' (i.e., no torque will be applied on the rotor by $s_2$). For simplicity, we do not consider backsteps in our simulations: each stator sees only the FliG in front of it. In time intervals where no stators are pushing on the rotor, stators are repositioned such that each is directly adjacent to a FliG.

All data points were computed as averages from 10 simulation runs, each of length 1 s. Because simulations were performed at low load, this run length was sufficient to include many steps. Standard errors of the mean were smaller than the size of markers.

## 3 Tether-wind calculation at high loads

In his 2003 review article, Howard Berg posed an argument for why torque-generating units in the flagellar motor must have a very high duty ratio [3]. This 'tether-wind' argument is summarized in the main text. Here, we redo this calculation with the values Berg originally used, and then revise it using our model construction and chosen parameters (given in Table S1). Though high-load simulations were not used in the conclusions for this manuscript, this section provides an explanation as to how high-load simulations were run in our previous paper [2].

Consider a cell tethered to a surface by its flagellar filament. The cell is spun around by the rotation of the motor at the base of the filament. In the first step of a resurrection experiment, a motor has a single torque-generating unit.

Berg estimated the torque generated by a wild-type motor (with 8 torque-generating units) to be 4000 pN-nm, so that each unit generates about 500 pN-nm of torque. Likewise, he estimated the torsional spring constant of the tether to be 500 pN-nm-rad$^{-1}$, leading to a twist in the tether of about 1 rad (57°). Since the cell body has a significantly higher drag than the rotor, the tether will unwind exponentially once the stator disengages: $\theta = \theta_0 \exp(-\alpha t)$, where $\theta_0$ is the initial twist and $\alpha$ is the torsional spring constant divided by the drag coefficient of the rotor. Estimating the drag of the rotor as 0.02 pN-nm-s-rad$^{-1}$, $\alpha = 2.5 \times 10^{-4}$ s$^{-1}$. Then, if the stator is disengaged for $1.6 \times 10^{-5}$ s (corresponding to a duty ratio of 0.999 in his calculation),

5

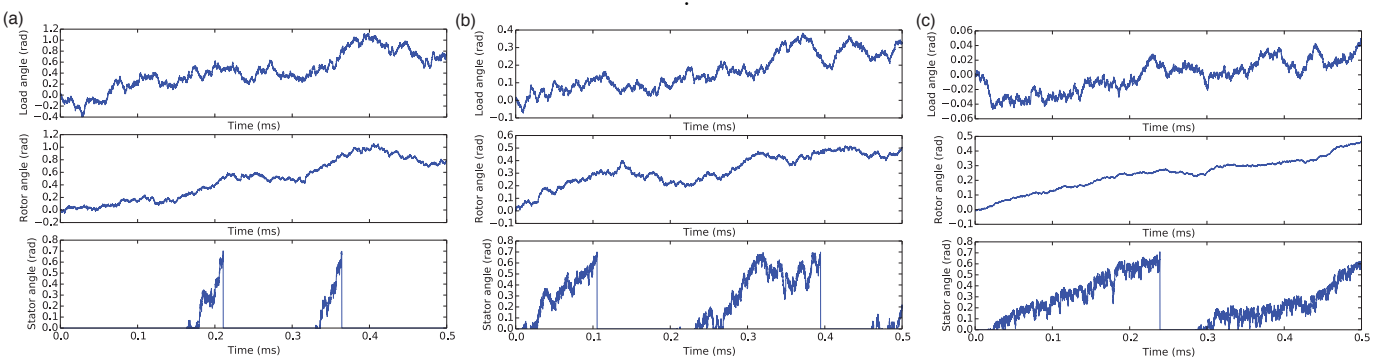

Figure S3: Simulation output (0.5 ms) for a single-stator motor at **(a)** $\zeta_L = 0.005$ pN-nm-s-rad$^{-1}$, **(b)** $\zeta_L = 0.05$ pN-nm-s-rad$^{-1}$, and **(c)** $\zeta_L = 0.5$ pN-nm-s-rad$^{-1}$. The duration of stator steps (bottom panel) increases with load, decreasing the relative amount of a mechanochemical cycle taken up by the waiting time between subsequent steps.

the twist in the tether decreases to $57\exp(-2.5 \times 10^4 \times 1.6 \times 10^{-5}) = 38°$, or by 19°. At the time of the publication, it was assumed that a single unit steps 50 times per revolution, so that a single step was approximately 7.2°, or less than half of the unwinding. This led to the conclusion that a single torque generator would not be able to keep up if it detached for a time even on the order of $10^{-5}$ s.

Most estimates have calculated the maximum torque in the BFM to be approximately 2000 pN-nm. Assuming a motor at stall has 11 stators, each stator generates approximately 180 pN-nm of torque. This is also consistent with single-stator measurements in chimeric motors [10]. We estimate the torsional spring constant very conservatively, at the lower end of the experimentally measured range, as 150 pN-nm-rad$^{-1}$ [6]. Then, the tether is twisted by 1.2 rad, or 69°.

The 'waiting time' between subsequent steps corresponds to the time required for an ion from the periplasm to bind to an exposed binding site on the stator. In our model simulations, this site is exposed when the angle of the stator $\phi_S < 0 + \epsilon$. Recall that $\langle T_w \rangle = 0.2$ ms. A stator disengages from the rotor from the time it completes its power stroke ($\phi_S \leq 0$) until an ion binds to it. In our simulations, when $\phi_S$ is in the interval $(0, 0 + \epsilon)$, it is able to bind a periplasmic cation while still being bound to the rotor. We take $\epsilon$ to be very small, $\frac{\pi}{1500}$ rad $= 0.12°$. For the vast majority of the loads considered, the time spent in this interval is negligible compared to $\langle T_w \rangle$, and the stator detaches from the rotor for 0.2 ms at a time, on average. However, in very slowly-rotating motors, the time when $\phi_S \in (0, 0 + \epsilon)$ may be large enough to significantly lower the average time that the stator detaches from the rotor.

Tethered cells rotated at 1.2 Hz, which corresponds to each step taking 32 ms (assuming there are 26 steps per revolution). Since the experiment is at very high load $\langle T_m \rangle \approx 32$ ms, since $\langle T_m \rangle \gg \langle T_w \rangle$. Because the BFM lives at low Reynolds number, we assume that the stator moves at a constant speed throughout its power stroke. Then $\phi_S \in (0, 0 + \epsilon)$ for $0.12°/20° = 0.006$ of $\langle T_m \rangle$, or 0.192 ms. Then, the average time that the stator is actually detached from the rotor between consecutive strokes at the load considered in our simulations is 0.008 ms. During this time, the tether unwinds to $69\exp(-150/0.02 \times 8 \times 10^{-6}) \approx 65°$, or by 4°. This is less than our assumed elementary step length, $2\pi/26 \approx 14°$.

However, we note here once again, as in the main text, that there is not yet concrete evidence that a single-stator motor at very high load does not, in fact, 'lose' several steps to the unwinding of the tether connection. This uncertainty will likely be resolved only by experiments which can quantify how the ion flux varies between single- and multi-stator motors (i.e., motors with different duty ratios) at high loads.

6



\end{document}